\documentstyle[axodraw]{article}

\title{Nucleon contribution to
the neutrino electromagnetic vertex in matter}
\author{
Juan Carlos D'Olivo \\
Instituto de Ciencias Nucleares\\
Universidad Nacional Aut\'{o}noma de M\'{e}xico\\
Apartado Postal 70-543, 04510 M\'{e}xico, D.F., M\'{e}xico
\and
Jos\'e F. Nieves\\
Laboratory of Theoretical Physics \\
Department of Physics \\
P.O. Box 23343, University of Puerto Rico\\
R\'{\i}o Piedras, Puerto Rico 00931-3343}
\date{hep-ph/9708391}

\begin{document}
\maketitle

\begin{abstract}

We calculate the nucleon contribution to the electromagnetic
vertex of a neutrino in a background of particles,
including the effect of the anomalous magnetic moment of the nucleons.
Explicit formulas for the form factors are given
in various physical limits of practical interest.
Several applications of the results are mentioned,
including the effect of an external magnetic field 
on the dispersion relation of a neutrino in matter.

\end{abstract}

\section{Introduction}

It is commonly accepted that the properties of neutrinos
that propagate through a thermal background of particles can be
very different compared to their properties in the vacuum.
This is the case, in particular, for their electromagnetic
properties\cite{np1}.
Some time ago, this notion motivated the study of 
the electromagnetic properties of neutrinos in a background of 
electrons\cite{dnp1}.  In Ref.\ \cite{dnp1}, the effective
electromagnetic interactions of a neutrino that propagates
through matter were determined by a one-loop calculation of
the electromagnetic vertex function induced by the neutrino
interactions with the electrons in the background.
The results of those calculations have been applied
to determine the rates of various physical processes such
as, for example, the radiative neutrino decay\cite{dnpradnudecay} 
and the Cherenkov radiation by massless (chiral) neutrinos\cite{dnpnucerenk},
the latter of which had been studied previously by different techniques\cite{nucerenk}.  We mention that the Cherenkov
process considered in Refs. \cite{dnpnucerenk,nucerenk} differs
from the one studied in the recent works by Grimus and Neufeld\cite{grimus}
and by Mohanty and Samal\cite{mohanty}, which depends on a hypothetical
magnetic moment coupling of the neutrino.

It was observed in Ref.\ \cite{dnp1} that, in the presence of a static
magnetic field, the induced electromagnetic interactions
produce an additional contribution to the effective 
potential of the neutrinos, or equivalently to their index of refraction,
that modify the Wolfenstein resonance condition\cite{wolfenstein}
for neutrino oscillations in matter.  This observation,
and the calculations on which it is based, have been generalized
and refined subsequently by various 
authors\cite{dn1,esposito,raffelt,smirnov}\footnote{A particularly 
clear exposition, which also corrects some
inaccurate statements contained in Ref.\ \cite{dn1}, has been given by
Smirnov in Ref.\ \cite{smirnovichep}}.

In the present work we extend the calculations carried out
in Ref.\ \cite{dnp1} by including the contribution to the
neutrino electromagnetic vertex comming from the presence
of the nucleons in the background.  
In particular, we take into account the anomalous magnetic
moment coupling of the nucleons to the photon and we
calculate explicitly the additional terms they induce
in the neutrino electromagnetic vertex.

The present calculation is motivated, in part, by the recent
interesting work of Kusenko and Segr\`e
suggesting that the observed large birth velocities of pulsars
are due to the asymmetric emission of neutrinos 
from the cooling protoneutron star, which
is produced by the resonant neutrino oscillations in the
supernova's magnetic field\cite{ks1}.  In a subsequent
paper\cite{ks2}, the same authors find that a similar
explanation is possible if the oscillations occur
between an active (weak interacting) neutrino and a sterile one.
While in the oscillation between active neutrinos the
neutral-current interaction contribution to the neutrino
energy in not relevant, it becomes important for oscillations
between an active and a sterile neutrino because the latter
has no weak interactions at all.  In these contexts, the effects
on the neutrino potentials due
to the magnetic couplings of the nucleons have been estimated
for various limiting cases in Refs. \cite{smirnov,sciama}.  

However, the calculations presented here go farther.  They are based
on the one-loop formula for the neutrino electromagnetic
vertex using thermal field theory methods.  Apart from the
limitations that the one-loop approximation (linear in the
electromagnetic field) imply, the formulas obtained for the contribution
due to the anomalous nucleon moments are valid for general 
conditions of the nucleon gas, degerate or non-degenerate,
whether it is relativistic or not.  The formulas can be applied,
in the context of neutrino oscillations
in the presence of a magnetic field,
to determine the additional corrections to the neutrino index
of refraction in situations in which the nucleons are not
necessarily described by one of the idealized limiting
cases, and instead a more detailed evaluation
of the effects is sought.  Besides the application
in this context, our calculations can be relevant for
other physical processes that have been considered in the
literature, such as the induced radiative neutrino decay
and the Cherenkov radiation emission by neutrinos mentioned above.

\section{Calculation of the neutrino electromagnetic vertex}
\label{sec:calculation}

\begin{figure}
\begin{center}
\begin{picture}(110,120)(0,0)
\Photon(30,60)(0,60){3}{5}
\Photon(90,60)(60,60){3}{5}
\ArrowLine(110,30)(90,60) 
\ArrowLine(90,60)(110,90)
\LongArrowArc(45,60)(15,90,450)
\Text(110,20)[]{$\nu(k)$}
\Text(110,100)[]{$\nu(k^\prime)$}
\Text(75,50)[]{$Z(q)$}
\Text(15,50)[]{$\gamma(q)$}
\Text(45,85)[]{$p,n$}
\end{picture}
\end{center}
\caption[]{One-loop diagram for the nucleon contributions
$\Gamma^{(p,n)}_\mu(k,k^\prime)$ to the neutrino electromagnetic 
vertex.}
\label{fig1}
\end{figure}
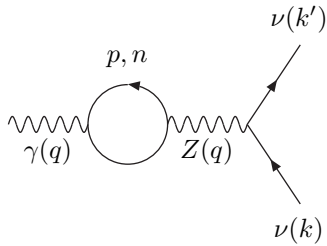

We follow the method and conventions of Ref.\ \cite{dn1}.  The 
background-dependent part of the neutrino electromagnetic
vertex function is denoted by $\Gamma^\prime_\mu(k,k^\prime,v)$
where $k$ and $k^\prime$ denote the momentum of the
incomming and outgoing neutrino, respectively,
and $v^\mu$ is the velocity four-vector of the medium,
which from now on we set it equal to $(1,\vec 0)$ in our
calculations.
The electron background contribution to $\Gamma^\prime_\mu$,
which we denote here by $\Gamma^{(e)}_\mu$,
was calculated in Ref.\ \cite{dn1} while the nucleon background
contribution $\Gamma^{(p)}_\mu$ + $\Gamma^{(n)}_\mu$ 
is the subject of the present work.  To lowest
order, the diagrams relevant to the calculation are 
shown in Fig.\ \ref{fig1}.
For each nucleon  $f = n,p$ in the loop, the propagator 
is given by
\begin{equation}\label{fprop}
S_F(p) = ({\not p} + m_f)
\left[\frac{1}{p^2 - m_f^2} + 2\pi i\delta(p^2 - m_f^2)\eta_f(p)
\right] \,,
\end{equation}
where
\begin{equation}\label{eta}
\eta_f(p) = \frac{\theta(p\cdot v)}{e^{\beta(p\cdot v - \mu_f)} + 1} +
\frac{\theta(-p\cdot v)}{e^{-\beta(p\cdot v - \mu_f)} + 1} \,,
\end{equation}
with $\theta$ representing the unit step function,
$1/\beta$ the temperatute and $\mu_f$ the chemical
potential of each nucleon specie.

The electromagnetic couplings of the nucleons
are given by
\begin{equation}\label{lgamma}
L_\gamma = -|e|A^\mu {\overline p}\gamma_\mu p
- \frac{\kappa_p}{2}\overline p\sigma^{\mu\nu}p F_{\mu\nu}
- \frac{\kappa_n}{2}\overline n\sigma^{\mu\nu}n F_{\mu\nu}\,,
\end{equation}
where $\kappa_{n,p}$ are the anomalous part of the nucleon
magnetic moments, given by
\begin{eqnarray}\label{kappanp}
\kappa_p & = & 1.79\left(\frac{|e|}{2m_p}\right)\,,\nonumber\\
\kappa_n & = & -1.91\left(\frac{|e|}{2m_n}\right)\,,\nonumber\\
\end{eqnarray}
$e$ stands for the electron charge
and, as usual, $\sigma_{\mu\nu} = \frac{i}{2}[\gamma_\mu,\gamma_\nu]$.
For the neutral-current couplings we write
\begin{equation}\label{lz}
L_Z = -g_Z Z^\mu\left[\overline\nu_L\gamma_\mu\nu_L + 
\sum_{f = e,p,n}\overline f\gamma_\mu(a_f + b_f\gamma_5)f\right]\,,
\end{equation}
where, in the standard model,
\begin{equation}
g_Z = g/(2\cos\theta_W)
\end{equation}
and
\begin{eqnarray}\label{neutralcouplings}
-a_e = a_p & = & \frac{1}{2} - 2\sin^2\theta_W \,,\nonumber\\
a_n & = & -\frac{1}{2}\,,\nonumber\\
b_e & = & \frac{1}{2}\,,\nonumber\\
b_n = -b_p & = & \frac{1}{2}g_A\,,
\end{eqnarray}
with $g_A = 1.26$ being the renormalization constant
of the axial-vector current of the nucleon.  There are
several implicit assumptions and simplifications that
we have made by adopting the electromagnetic
and neutral-current couplings defined by Eqs.\ (\ref{lgamma}) and (\ref{lz}).
Their justification is discussed in more detail
in the appendix.

With these couplings, the nucleon contribution 
to the neutrino electromagnetic
vertex is given by
\begin{equation}\label{gammaf}
\Gamma^{(nucl)}_{\mu} = (T^{(p)}_{\mu\nu} + T^{(n)}_{\mu\nu})\gamma^\nu L\,,
\end{equation}
where, as usual $L = (1 - \gamma_5)/2$, and
\begin{eqnarray}\label{tmunu}
T_{\mu\nu}^{(f)} 
& = & \sqrt{2}G_F
\int\frac{d^4p}{(2\pi)^3}\mbox{\bf Tr}
\left[j^{(em)}_{f\mu}(q)(\not p + m_f)
\gamma_\nu(a_f + b_f \gamma_5)
(\not p - \not q + m_f)\right]\times\nonumber\\
& & \left\{
\frac{\delta(p^2 - m_f^2)\eta_f(p)}{(p - q)^2 - m_f^2} +
\frac{\delta[(p - q)^2 - m_f^2]\eta_f(p - q)}{p^2 - m_f^2}
\right\}\,,
\end{eqnarray}
for $f = n,p$.
To arrive at this formula
we have dropped the term
that is independent of the particle density distributions,
as well as the term that contains the product of the
two delta functions.\footnote{The latter contributes only
to the absorptive part (provided the appropriate kinematical
conditions are satisfied), which we are not considering
in this paper.}
In Eq.\ (\ref{tmunu}) $q = k - k^\prime$ denotes the momentum of 
the outgoing photon, $j^{(em)}_{f\mu}(q)$ is the total
electromagnetic current of each nucleon,
\begin{eqnarray}\label{jnucl}
j^{(em)}_{p\mu}(q) & = & |e|\gamma_\mu -i\kappa_p\sigma_{\mu\alpha}q^\alpha\nonumber\\
j^{(em)}_{n\mu}(q) & = & -i\kappa_n\sigma_{\mu\alpha}q^\alpha\,,
\end{eqnarray}
and we have also used the relation
\begin{equation}
\frac{g_Z^2}{m_Z^2} = \sqrt{2}G_F\,.
\end{equation}
Let us consider the neutron case first.
Making the change of variable $p\rightarrow p + q$ in the 
integrand corresponding to the second
term in brackets in Eq.\ (\ref{tmunu}),
and carrying out the traces, we obtain\footnote{Our conventions
are such that $\epsilon^{0123} = +1$.}
\begin{eqnarray}\label{tmunun}
T_{\mu\nu}^{(n)} & = & 4m_n\kappa_n\sqrt{2}G_F
\int\frac{d^3{\cal P}}{(2\pi)^3 2{\cal E}}
\left\{a_{n}(q^2 g_{\mu\nu} - q_\mu q_\nu)(f_n + f_{\overline n})
\right.\nonumber\\
& & \left. - 2ib_{n}\epsilon_{\mu\nu\alpha\beta}q^\alpha 
p^\beta(f_n - f_{\overline n})
\right\}
\times\left[\frac{1}{q^2 + 2p\cdot q} + (q\rightarrow -q)\right]\,,
\end{eqnarray}
where
\begin{equation}\label{momentum}
p^\mu = ({\cal E},\vec {\cal P}),\quad {\cal E} = 
\sqrt{\vec {\cal P}^{\, 2} + m_n^2}\,.
\end{equation}
We have introduced the number densities of the
nucleons
\begin{equation}\label{fnp}
f_{n,p}(p) = \frac{1}{e^{\beta({\cal E} - \mu_{n,p})} + 1}\,,
\end{equation}
and the corresponding quantities $f_{\overline n,\overline p}$ for
the antiparticles, which are given by a similar formula
but with the opposite sign of the chemical potential.
Following Ref.\ \cite{dnp1}, $T_{\mu\nu}^{(n)}$ can be decomposed
in the form
\begin{equation}\label{tmunundecomp} 
T^{(n)}_{\mu\nu} = T^{(n)}_T R_{\mu\nu} + T^{(n)}_L Q_{\mu\nu}
+ T^{(n)}_P P_{\mu\nu}\,,
\end{equation}
where
\begin{eqnarray}\label{tensors}
R_{\mu\nu} & = & \tilde g_{\mu\nu} - Q_{\mu\nu}\,,\nonumber\\
Q_{\mu\nu} & = & \frac{\tilde v_\mu\tilde v_\nu}{{\tilde v}^2}\,\nonumber\\
P_{\mu\nu} & = & \frac{i}{\cal Q}\epsilon_{\mu\nu\alpha\beta}q^\alpha v^\beta\,,
\end{eqnarray}
with
\begin{equation}\label{gtilde}
\tilde g_{\mu\nu} = g_{\mu\nu} - \frac{q_\mu q_\nu}{q^2}
\end{equation}
and
\begin{equation}\label{vtilde}
\tilde v_\mu = \tilde g_{\mu\nu} v^\nu\,.
\end{equation}
The functions $T^{(n)}_{T,L,P}$ depend on the variables
\begin{eqnarray}\label{photonvars}
\Omega & = & q\cdot v\nonumber\\
{\cal Q} & = & \sqrt{\Omega^2 - q^2}\,,
\end{eqnarray}
which are the energy and momentum of the photon,
and they are given explicitly by
\begin{eqnarray}\label{Tneutron}
T^{(n)}_T = T^{(n)}_L & = & 4\sqrt{2}G_Fm_n \kappa_n a_n q^2 D_n\nonumber\\
T^{(n)}_P & = & -8\sqrt{2}G_F m_n \kappa_n b_n {\cal Q}C_n\,,
\end{eqnarray}
where
\begin{eqnarray}\label{D}
D_{n,p} & = & \int\frac{d^3{\cal P}}{(2\pi)^3 2{\cal E}}
(f_{n,p} + f_{\overline n,\overline p})\left[
\frac{1}{q^2 + 2p\cdot q} + (q\rightarrow -q)\right]\\
\label{C}
C_{n,p} & = & \int\frac{d^3{\cal P}}{(2\pi)^3 2{\cal E}}
\left(\frac{\tilde v\cdot p}
{{\tilde v}^2}\right)
(f_{n,p} - f_{\overline n,\overline p})\left[
\frac{1}{q^2 + 2p\cdot q} + (q\rightarrow -q)\right]\,.
\end{eqnarray}
It is useful to recall that
\begin{eqnarray}
{\tilde v}^2 & = & -\frac{{\cal Q}^2}{q^2}\,\nonumber\\
\tilde v\cdot p & = & {\cal E} - \left(\frac{q\cdot p}{q^2}\right)\Omega\,.
\end{eqnarray}

According to Eq.\ (\ref{jnucl}), the proton contribution
to the $\nu\nu\gamma$ vertex contains a term that is similar to the
one determined above in the neutron case, plus
another one that arises from the ordinary
$\gamma_\mu$ coupling.  The latter
is of the same of form as the one calculated
in Ref.\ \cite{dnp1} for the neutral-current contribution
in the electron background.  Thus, repeating the steps
that lead to Eq. (2.24) of Ref.\ \cite{dnp1} and imitating
Eq.\ (\ref{Tneutron}) above, we obtain the total contribution
from the proton background
\begin{eqnarray}\label{Tprot}
T^{(p)}_T & = & 2\sqrt{2}G_F a_p\left\{
|e|\left(A_{p} - \frac{B_{p}}{{\tilde v}^2}\right) + 
2m_p \kappa_p q^2 D_{p}\right\}\nonumber\\
T^{(p)}_L & = & 4\sqrt{2}G_F a_p\left\{
|e|\frac{B_{p}}{{\tilde v}^2} +
m_p \kappa_p q^2 D_{p}\right\}\nonumber\\
T^{(p)}_P & = & -4\sqrt{2}G_F b_p {\cal Q}C_{p}\left(
|e| + 
2 m_p \kappa_p\right)\,,
\end{eqnarray}
where
\begin{eqnarray}\label{AB}
A_p & = & \int\frac{d^3{\cal P}}{(2\pi)^3 2{\cal E}}
(f_{p} + f_{\overline p})\left[
\frac{2m_p^2 - 2p\cdot q}{q^2 + 2p\cdot q} + (q\rightarrow -q)\right]\nonumber\\
B_p & = & \int\frac{d^3{\cal P}}{(2\pi)^3 2{\cal E}}
(f_{p} + f_{\overline p})\nonumber\\
& & \times\left[
\frac{2(p\cdot v)^2 + 2(p\cdot v)(q\cdot v) - p\cdot q}
{q^2 + 2p\cdot q} + (q\rightarrow -q)\right]\,.
\end{eqnarray}

The expressions in Eqs.\ (\ref{C}), (\ref{D}) and 
(\ref{AB}) allow us to obtain
simple formulas for the coefficients
$T^{(n,p)}_{T,L,P}$ in various limiting cases.
For Eqs.\ (\ref{C}) and (\ref{AB}) we can borrow the results from
Ref.\ \cite{dnp1}, where the corresponding quantities for
the electron background were denoted by $A,B,C$.
Thus, for example, for the proton background
\footnote{We take the opportunity
to mention the following typographical errors in
Ref.\ \cite{dnp1}: the second formula
for $C$ in Eq. (2.24) of Ref.\ \cite{dnp1} contains the factor
$f_{-} + f_{+}$ when it should be $f_{-} - f_{+}$, and 
the left-hand side of Eq. (2.28) of the same reference
should be $B$ instead
of $B/{\tilde v^2}$.  The
electron and positron distribution functions are denoted
by $f_{e,\overline e}$ in the present paper.},
\begin{eqnarray}\label{Aex}
A_p(\Omega,{\cal Q}\rightarrow 0) & = & -3\omega^2_{0p} + 
\frac{{\cal Q}^2 \omega^2_{0p}}{\Omega^2} +
O(\Omega^2)\,,\nonumber\\
A_p(\Omega\rightarrow 0,{\cal Q}) & = & 
\frac{1}{2}\int\frac{d^3{\cal P}}{(2\pi)^3} \frac{d}{d{\cal E}}
(f_p + f_{\overline p}) + O({\cal Q}^2)\,,
\end{eqnarray}
\begin{eqnarray}\label{Bex}
B_p(\Omega,{\cal Q}\rightarrow 0) & = &
\frac{{\cal Q}^2 \omega^2_{0p}}{\Omega^2}\,,\nonumber\\
B_p(\Omega\rightarrow 0,{\cal Q}) & = &
A_p(\Omega\rightarrow 0,{\cal Q}) + O({\cal Q}^2)\,,
\end{eqnarray}
\begin{eqnarray}\label{Cex}
C_p(\Omega,{\cal Q}\rightarrow 0) & = &
-\frac{1}{2}\int\frac{d^3{\cal P}}{(2\pi)^3 2{\cal E}} 
\frac{f_p - f_{\overline p}}{{\cal E}}\left[1 -
\frac{2{\cal P}^2}{3{\cal E}^2}\right] + O(\Omega^2)\,,
\nonumber\\
C_p(\Omega\rightarrow 0,{\cal Q}) & = &
\frac{1}{2}\int\frac{d^3{\cal P}}{(2\pi)^3 2{\cal E}}
\frac{d}{d{\cal E}}
(f_p - f_{\overline p}) + O({\cal Q}^2)\,,
\end{eqnarray}
where
\begin{equation}\label{Omega0}
\omega^2_{0p} = \int\frac{d^3{\cal P}}{(2\pi)^3 2{\cal E}}
(f_p + f_{\overline p})\left[1 - \frac{{\cal P}^2}{3{\cal E}^2}
\right]\,.
\end{equation}
In similar fashion, we obtain here
\begin{eqnarray}\label{Dex}
D_p(\Omega,{\cal Q}\rightarrow 0) & = &
-\frac{1}{2}\int\frac{d^3{\cal P}}{(2\pi)^3 2{\cal E}} 
\frac{f_p + f_{\overline p}}{{\cal E}^2} + O(\Omega^2)\,,\nonumber\\
D_p(\Omega\rightarrow 0,{\cal Q}) & = &
\frac{1}{2}\int\frac{d^3{\cal P}}{(2\pi)^3 2{\cal E}}
\frac{d}{d{\cal E}}
\left(\frac{f_p + f_{\overline p}}{{\cal E}}\right) + O({\cal Q}^2)\,.
\end{eqnarray}
We would like to stress that, while the previous formulas hold
for the limiting values indicated of the photon energy
and momentum,
no assumption has been made with respect to the background
gas.  The formulas can be simplified further by restricting
the attention to some simple idealized situations,
such as the degenerate and non-degenerate cases, in both
the relativistic and non-relativistic limits.
Since the nucleons are non-relativistic in the situations 
of practical interest, we consider this particular case 
in some detail.  

In the non-relativistic limit, the above expressions
reduce to
\begin{equation}\label{Omega0nr}
\omega^2_{0p} = \frac{n_p}{4m_p}\,,
\end{equation}
\begin{equation}\label{Aex2}
A_p(\Omega\rightarrow 0,{\cal Q}) =
\frac{-m_p}{4\pi^2}I_p + O({\cal Q}^2)\,, 
\end{equation}
\begin{eqnarray}\label{Cex2}
C_p(\Omega,{\cal Q}\rightarrow 0) & = &
\frac{-\omega^2_{0p}}{2m_p} + O(\Omega^2)\,,\nonumber\\
C_p(\Omega\rightarrow 0,{\cal Q}) & = &
\frac{-1}{8\pi^2}I_p + O({\cal Q}^2)\,, 
\end{eqnarray}
\begin{eqnarray}\label{Dex2}
D_p(\Omega,{\cal Q}\rightarrow 0) & = &
\frac{-\omega^2_{0p}}{2m^2_p} + O(\Omega^2)\,,\nonumber\\
D_p(\Omega\rightarrow 0,{\cal Q}) & = &
\frac{-1}{8\pi^2 m_p}I_p + O({\cal Q}^2)\,,
\end{eqnarray}
where $n_p$ is the total proton number density 
\begin{equation}\label{np}
n_p = 2\int \frac{d^3{\cal P}}{(2\pi)^3}f_p\,,
\end{equation}
and
\begin{equation}\label{Int}
I_p = \int_0^\infty d{\cal P}f_p \,.
\end{equation}
The integral in Eq.\ (\ref{Int}) cannot be evaluated without
knowing the distribution function and, therefore, it
depends on whether the background is degenerate
or non-degenerate.  For these two limiting cases
we obtain
\begin{equation}\label{Intlimits}
I_p =
\left\{
\begin{array}{ll}
\pi^2\frac{\beta n_p}{m_p}
& \mbox{(non-degenerate)}\,,\\
{}\\
(3\pi^2 n_p)^{1/3}
& \mbox{(degenerate)}\,.
\end{array}
\right.
\end{equation}
The corresponding formulas for the neutron background 
are easily obtained from the above by making
obvious substitutions.  In this manner, via
Eqs.\ (\ref{Tneutron}) and (\ref{Tprot}), we determine
the nucleon contribution to the neutrino electromagnetic
vertex either in the static ($\Omega\rightarrow 0$)
or in the long wavelength (${\cal Q}\rightarrow 0$) limit.
It is useful for some applications
to consider the case in which neither $\Omega$ nor ${\cal Q}$
is zero, while still satisfying $\Omega, {\cal Q}\ll m_{n,p}$,
which is a good approximation for most situations of interest.
The results for this case are derived in Appendix \ref{app:smallq}.

\section{Discussion and Applications}

The total matter background contribution to the $\nu\nu\gamma$
vertex is given by
\begin{equation}\label{Gammamatt}
\Gamma_\mu = (T_T R_{\mu\nu} + T_L Q_{\mu\nu} + 
T_P P_{\mu\nu})\gamma^\nu L\,,
\end{equation}
where
\begin{equation}\label{Ttot}
T_X = T^{(p)}_X + T^{(n)}_X + T^{(e)}_X \quad (X = T,L,P)\,.
\end{equation}
The electron term $T^{(e)}_{\mu\nu}$ can be decomposed
as in Eq.\ (\ref{tmunundecomp}), with
\begin{eqnarray}\label{Te}
T^{(e)}_T & = & 2\sqrt{2}eG_F\left(A - \frac{B}{{\tilde v}^2}\right)
\left\{\begin{array}{ll}
a_e + 1 & (\nu_e)\\
{}\\
a_e & (\nu_{\mu,\tau})
\end{array}
\right. \nonumber\\
T^{(e)}_L & = & 4\sqrt{2}eG_F\frac{B}{{\tilde v}^2}
\left\{\begin{array}{ll}
a_e + 1 & (\nu_e)\\
{}\\
a_e & (\nu_{\mu,\tau})
\end{array}
\right. \nonumber\\
T^{(e)}_P & = & -4\sqrt{2}eG_F{\cal Q}C
\left\{\begin{array}{ll}
b_e - 1 & (\nu_e)\\
{}\\
b_e & (\nu_{\mu,\tau})
\end{array}
\right. 
\end{eqnarray}
The additional contribution for the electron neutrino
is due to the 
charged-current diagram, which is absent for $\nu_{\mu,\tau}$.
The functions $A,B,C$ are given by expressions analogous
to those
in Eqs.\ (\ref{C}) and (\ref{AB}), with $f_{n,p}$ being replaced by the electron
distribution $f_e$.  In the non-relativistic limit they reduce
to formulas analogous to those given in Eqs.\ (\ref{Aex2}) and (\ref{Cex2}).

Since the electrons are relativistic in many situations
of interest, it is also useful to summarize the corresponding results
in that limit.  Thus, for a relativistic electron gas,
\begin{equation}\label{omega0ur}
\omega^2_0 = \frac{1}{6\pi^2}\int_0^\infty 
d{\cal P}\,{\cal P}(f_e + f_{\overline e})\,,
\end{equation}
and
\begin{eqnarray}\label{ACeur}
A(0,{\cal Q}) & = & -3\omega^2_0 + O({\cal Q}^2)\,,\nonumber\\
C(0,{\cal Q}) & = & -\frac{1}{8\pi^2}(I_e - I_{\overline e})
+ O({\cal Q}^2)\,,\nonumber\\
C(\Omega,0) & = & -\frac{1}{24\pi^2}(I_e - I_{\overline e})
+ O(\Omega^2)\,,
\end{eqnarray}
where
\begin{equation}\label{Je}
I_{e,\overline e} = \int_0^\infty d{\cal P}\,f_{e,\overline e}\,.
\end{equation}
The remaining integrals in Eqs.\ (\ref{omega0ur}) and (\ref{Je}) cannot be performed
without specifying the distribution function.  In the limiting
cases of a degenerate or non-degenerate gas, they are given by
\begin{eqnarray}\label{omega0urex}
\omega^2_0 & = & 
\left\{
\begin{array}{ll}
\frac{\beta}{12}(n_e + n_{\overline e})
& \mbox{(non-degenerate)}\,,\\
{}\\
\frac{1}{3}\left(\frac{3}{8\pi}\right)^{2/3}(n^{2/3}_e + n^{2/3}_{\overline e})
& \mbox{(degenerate)}\,,
\end{array}
\right. \\
{}\nonumber\\
\label{Jeex}
I_{e,\overline e} & = & 
\left\{
\begin{array}{ll}
\frac{\pi^2\beta^2}{2}n_{e,\overline e}
& \mbox{(non-degenerate)}\,,\\
{}\\
(3\pi^2 n_{e,\overline e})^{1/3}
& \mbox{(degenerate)}\,.
\end{array}
\right.
\end{eqnarray}
It should also be remembered that, for either case,
\begin{eqnarray}\label{ABeany}
A(\Omega,{\cal Q}\rightarrow 0) & = & -3\omega^2_0 + 
\frac{{\cal Q}^2 \omega^2_{0}}{\Omega^2} +
O(\Omega^2)\,,\nonumber\\
B(\Omega,{\cal Q}\rightarrow 0) & = & 
\frac{{\cal Q}^2\omega^2_0}{\Omega^2}\,,\nonumber\\
B(\Omega\rightarrow 0,{\cal Q}) & = &
A(\Omega\rightarrow 0,{\cal Q}) + O({\cal Q}^2)\,.
\end{eqnarray}
The application of these and the formulas obtained in Section \ref{sec:calculation}
depend on the specific environment under consideration as well
as the kinematic regime involved.  Let us then consider some
particular situations of interest.

From the explicit formulas given in Eq.\ (\ref{Dex2}), or more generally
in Eq.\ (\ref{Dsmallq}), it is immediately seen that for values
of $q$ such that $\Omega, {\cal Q}\ll m_{n,p}$, the function
$D_{n,p}$ is smaller than $A_p$ and $B_p$ by a factor of order
$1/m_{n,p}^2$.  Therefore, neglecting such terms in
Eqs.\ (\ref{Tneutron}) and (\ref{Tprot})
it follows that Eq.\ (\ref{Ttot}) reduces to 
\begin{eqnarray}\label{Ttotapprox1}
T_T & = & 2\sqrt{2}|e|G_F a_p\left(A_p - \frac{B_p}{{\tilde v}^2}\right)
+ T_T^{(e)}\nonumber\\
T_L & = & 4\sqrt{2}|e|G_F a_p \frac{B_p}{{\tilde v}^2}
+ T_L^{(e)}\,,
\end{eqnarray}
with the $T_{T,L}^{(e)}$ given in Eq.\ (\ref{Te}).  The relative
importance of the electron and the proton contributions
in these formulas depend on the kinematic regime involved
as well as the conditions of the proton and electron gases.

For illustrative purposes, suppose that the physical situation
is such that $q$ satisfies 
\begin{equation}\label{smallerq}
\Omega, {\cal Q} \ll m_e\,,
\end{equation}
in which case Eq.\ (\ref{smallq}) is satisfied also.  If both
the electron and proton gases are non-degenerate and
non-relativistic, then from Eqs.\ (\ref{Aex}) and (\ref{Omega0nr})
and the corresponding formulas for the electron we have
\begin{eqnarray}\label{ex1}
A_p(\Omega,0) & = & -\frac{3n_p}{4m_p}\nonumber\\
A(\Omega,0) & = & -\frac{3n_e}{4m_e}\,,
\end{eqnarray}
with analogous results for $B_p$ and $B$.
Thus in this case the proton contribution to $T_{T,L}(\Omega,0)$
is neglegible.  On the other hand, Eqs.\ (\ref{Aex2}) and (\ref{Intlimits}),
and the analogous formulas for the electron, imply
\begin{eqnarray}\label{ex2}
B_p(0,{\cal Q}) = A_p(0,{\cal Q}) & = & -\frac{1}{4}\beta n_p\nonumber\\
B(0,{\cal Q}) = A(0,{\cal Q}) & = & -\frac{1}{4}\beta n_e\,,
\end{eqnarray}
so that the proton and the electron contributions to 
$T_{T,L}(0,{\cal Q})$ are comparable.  This last conclusion
remains valid even if the electrons are relativistic.  In fact,
in that case, their contribution is also given
by the result given in Eq.\ (\ref{ex2}), as can be easily
checked using Eq.\ (\ref{omega0urex}) in Eqs.\ (\ref{ABeany}) and (\ref{ACeur}).
More possibilities
can obviously arise if we consider other realistic
situations, such as a nondegenerate proton gas but
a degenerate electron gas, or a kinematic regime
in which $\Omega, {\cal Q}\ll m_p$ is still satisfied
but Eq.\ (\ref{smallerq}) is not.

\section{Neutrino dispersion relation in a magnetic field}\label{sec:numag}

In the presence of a static, uniform magnetic field $\vec B$,
the $\nu\nu\gamma$ modifies the neutrino
dispersion relation in the medium according to
\begin{equation}\label{disprel}
\omega_k = |\vec k| + b - c\hat k\cdot\vec B\,,
\end{equation}
where $\vec k$ is the momentum vector of the neutrino
and $b$ gives the standard Wolfenstein term in the dispersion
relation\cite{wolfenstein,bterm}.  As shown in Ref.\ \cite{dnp1},
\begin{equation}\label{cterm}
c = \left[\frac{T_P(0,{\cal Q}}{\cal Q}\right]_{{\cal Q}
\rightarrow 0}\,.
\end{equation}
Substituting the formulas for $T^{(n,p,e)}_P$ given
in Eqs.\ (\ref{Tneutron}), (\ref{Tprot}) and (\ref{Te}),
this yields
\begin{eqnarray}\label{cterm2}
c & = & -4\sqrt{2}G_F\left[\phantom{\frac{1}{2}}
b_p(|e| + 2m_p\kappa_p)C_p(0,{\cal Q}\rightarrow 0)\right.\nonumber\\
& & \left.\mbox{} + b_n 2m_p\kappa_n C_n(0,{\cal Q}\rightarrow 0) \mp \frac{1}{2}
eC(0,{\cal Q}\rightarrow 0)\right]\,,
\end{eqnarray}
where the upper (lower) sign holds for $\nu_e$ ($\nu_{\mu\tau}$)
and we have put $b_e = \frac{1}{2}$.  If the electron gas is
degenerate, then
\begin{equation}\label{Cedeg}
C(0,{\cal Q}\rightarrow 0) = -\frac{1}{8}\left(
\frac{3n_e}{\pi^4}\right)^{1/3}\,,
\end{equation}
for both the relativistic and non-relativistic cases.
If the physical situation is such that the proton
gas also is degenerate, then a similar formula holds
for $C_p(0,{\cal Q}\rightarrow 0)$ (with $n_e\rightarrow n_p$)
and, in a neutral system, the electron and the normal
proton contributions tend to cancel for $\nu_e$ in Eq.\ (\ref{cterm2}).
In fact, if the effect of the anomalous nucleon magnetic
moment as well as  the renormalization of the nucleon
axial coupling are neglected, then $c$ in Eq.\ (\ref{cterm2}) would
be zero for $\nu_e$.
However, the cancellation is not complete once those two
effects are taken into account, independently of whether
the proton gas is degenerate or non-degenerate.  This 
can have important consequences in the context of the
possible explanation of the pulsar birth velocicites in terms
of resonant oscillations between active and sterile neutrinos\cite{ks2},
as recently pointed out in Ref.\ \cite{smirnov,sciama}.  Eq.\ (\ref{cterm2}),
together with the formulas for $C_{p,n}(0,{\cal Q})$ and
$C(0,{\cal Q})$ given in Section \ref{sec:calculation}
[e.g., Eq.\ (\ref{Cex}) and the corresponding formulas for $C_n$ and $C$]
give the magnetic contribution to the neutrino
dispersion relation for fairly general conditions
of the matter background.  Under some circumstances, it may
be more appropriate to use other methods to determine
this contribution, such as those employed in 
Refs. \cite{raffelt,smirnov,smirnovichep}.

\section{Conclusions}

In this work we have extended the previous calculations
of the electromagnetic properties of neutrinos
in a background of electrons, by including the
contribution from the nucleon background.  In particular,
we have taken into account the anomalous electromagnetic
and neutral-current couplings of the nucleons. 
The calculations are based
on the one-loop formula for the neutrino electromagnetic
vertex using thermal field theory methods and, apart from the
limitations of the one-loop approximation,
the formulas obtained for the electromagnetic vertex 
are valid for general 
conditions of the nucleon gas, degerate or non-degenerate,
whether it is relativistic or not.  
In the context of neutrino oscillations
in the presence of a magnetic field,
we applied the formulas
to determine the additional corrections to the neutrino index
of refraction in those situations in which the nucleons are not
necessarily described by one of the idealized limiting
cases, and instead a more detailed evaluation
of the effects is sought.  We have already mentioned
that the formulas for the electromagnetic vertex have been
the basis for studying other physical processes of neutrinos,
such as the induced radiative decay and Cherenkov radiation.
Here we have shown that
the importance of the nucleon contribution to
the functions $T_{T,L,P}$, relative
to the electron contribution, depends on the particular
physical conditions of the situation under consideration.
For example, we indicated how different the results can be
depending on whether
the nucleon and electron gases are degenerate or not,
or whether the kinematic regime
is such that $\Omega, {\cal Q}\gg m_e$
or $\Omega, {\cal Q}\ll m_e$ while still maintaining
$\Omega, {\cal Q}\ll m_p$, among other
possibilities.  The formulas that we have given in
this paper form a useful starting point to
study in more detail the radiative neutrino process
in such astrophysical settings
as the supenova and the early universe,
or perhaps in the context of laboratory experiments
involving the coherent neutrino electromagnetic conversion in
crystals that has been discussed recently\cite{zoller}.

This work has been partially supported by the U.S. National
Science Foundation Grant PHY-9600924 (JFN) and by
Grant No. DGAPA-IN100694 at the Universidad Nacional
Aut\'onoma de M\'exico (JCD).

\appendix
\section{The limit of small photon momentum}\label{app:smallq}

Here we consider the functions $A_f, B_f, C_f, D_f$
for values of the photon momentum satisfying
\begin{equation}\label{smallq}
\Omega, {\cal Q} \ll m_{p,n}\,,
\end{equation}
but for in the case that neither $\Omega$ nor ${\cal Q}$
is necessarily zero.  Consider first the formula for $B_p$ given
in Eq.\ (\ref{AB}).  It can be written as
\begin{equation}
B_p = \frac{1}{2}\int\frac{d^3{\cal P}}{(2\pi)^3}
(f_{p} + f_{\overline p})\left[
\frac{2{\cal E} + 2\Omega + \vec v_{\cal P}\cdot\vec{\cal Q}}
{q^2 + 2p\cdot q} + (q\rightarrow -q)\right]\,,
\end{equation}
where
$\vec v_{\cal P}\equiv \vec{\cal P}/{\cal E}$ is the velocity
of the background particles.  
We now make the change of variable
\begin{equation}
\vec{\cal P}\rightarrow \vec{\cal P} - \frac{1}{2}\vec{\cal Q}\,.
\end{equation}
After expanding the numerator and denominator in the terms
inside the brackets and neglecting terms of order
${\cal Q}^2/{\cal E}^2$, 
\begin{equation}
B_p = \frac{1}{2}\int\frac{d^3{\cal P}}{(2\pi)^3}\left[
\frac{(f_{p}(\vec{\cal P} - \frac{1}{2}\vec{\cal Q}) + 
f_{\overline p}(\vec{\cal P} - \frac{1}{2}\vec{\cal Q})}
{\Omega - \vec v_{\cal P}\cdot\vec {\cal Q}}
+ (q\rightarrow -q)\right]\,.
\end{equation}
Finally, expanding the distribution
functions in powers of $\vec {\cal Q}$, we obtain
\begin{equation}\label{Bsmallqa}
B_p = -\frac{1}{2}\int\frac{d^3{\cal P}}{(2\pi)^3}
\frac{\vec{\cal Q}\cdot\nabla_{\cal P}(f_p + f_{\overline p})}
{\Omega - \vec v_{\cal P}\cdot\vec{\cal Q}}\,,
\end{equation}
where $\nabla_{\cal P}$ is the gradient operator with respect
to the momentum variable $\vec{\cal P}$.
Since the distribution functions depend on
${\cal P}$ only through ${\cal E}$, Eq.\ (\ref{Bsmallqa})
is equivalent to
\begin{equation}\label{Bsmallqb}
B_p = -\frac{1}{2}\int\frac{d^3{\cal P}}{(2\pi)^3}
\left(\frac{\vec v_{\cal P}\cdot\vec{\cal Q}}{\Omega - \vec v_{\cal P}
\cdot\vec{\cal Q}}
\right)\frac{d}{d{\cal E}}(f_p + f_{\overline p})\,.
\end{equation}

In the limit $\Omega\rightarrow 0$, 
this formula reduces
to the result quoted in Eq.\ (\ref{Bex}).  It is also straightforward
to show, after performing an integration by parts, that
it reproduces the result quoted in Eq.\ (\ref{Bex}) for the
limit ${\cal Q}\rightarrow 0$.  However, it is important
to remark that
Eq.\ (\ref{Bsmallqb}) holds for any arbitrary values of $\Omega, {\cal Q}$,
subject only to the condition in Eq.\ (\ref{smallq}).  

We can proceed in similar fashion with the function $A_p$,
although the algebra is somewhat more involved in this case.
Thus, making the change of variable 
$\vec{\cal P}\rightarrow \vec{\cal P} - \frac{1}{2}\vec{\cal Q}$
in the formula for $A_p$ given in Eq.\ (\ref{AB}), then expanding the
integrand in powers of ${\cal Q}/E$ and neglecting
terms of order ${\cal Q}^2/E$ we obtain
\begin{equation}
\label{Asmallqa}
A_p = B_p + \int\frac{d^3{\cal P}}{(2\pi)^3}
\left[\frac{
\vec{\cal Q}\cdot\nabla_{\cal P}(f_p + f_{\overline p})
(\vec{\cal P}\cdot\vec v_{\cal P})
}
{2({\cal E}\Omega - \vec{\cal P}\cdot\vec{\cal Q})}
- \frac{(3 - v_{\cal P}^2)}{2{\cal E}}{(f_p + f_{\overline p})}
\right]\,.
\end{equation}
With the help of the identity
\begin{eqnarray}
(\vec{\cal P}\cdot\vec v_{\cal P})
\vec{\cal Q}\cdot\nabla_{\cal P} & = & 
(\vec{\cal P}\cdot\vec{\cal Q})
\vec v_{\cal P}\cdot\nabla_{\cal P}\nonumber\\
& = & 
[{\cal E}\Omega - ({\cal E}\Omega - \vec{\cal P}\cdot\vec{\cal Q})]
\vec v_{\cal P}\cdot\nabla_{\cal P}\,,
\end{eqnarray}
Eq.\ (\ref{Asmallqa}) reduces to
\begin{equation}
\label{Asmallqb}
A_p = B_p + \frac{\Omega}{2}\int\frac{d^3{\cal P}}{(2\pi)^3}
\frac{\vec v_{\cal P}\cdot\nabla_{\cal P}(f_p + f_{\overline p})}
{\Omega - \vec v_{\cal P}\cdot\vec{\cal Q}}\,.
\end{equation}
In Eq.\ (\ref{Asmallqb}) we have omitted the terms
\begin{equation}\label{omitted}
\int\frac{d^3{\cal P}}{(2\pi)^3}
\left[-\frac{1}{2}\vec v_{\cal P}\cdot\nabla_{\cal P}(f_p + f_{\overline p})
- \frac{(3 - v_{\cal P}^2)}{2{\cal E}}(f_p + f_{\overline p})
\right]
\end{equation}
which, using the fact that 
\begin{displaymath}
\nabla_{\cal P}\cdot\vec v_{\cal P} = \frac{3 - v_{\cal P}^2}{{\cal E}}
\end{displaymath}
reduce (apart from a factor of -1/2) to the integral of 
$\nabla_{\cal P}\cdot[\vec v_{\cal P}(f_p + f_{\overline p})]$
and therefore integrate to zero.  Eq.\ (\ref{Asmallqb}) can be written
in the equivalent form
\begin{equation}\label{Asmallqc}
A_p = B_p + \frac{\Omega}{2}\int\frac{d^3{\cal P}}{(2\pi)^3}
\left(
\frac{v_{\cal P}^2}
{\Omega - \vec v_{\cal P}\cdot\vec{\cal Q}}
\right)
\frac{d}{d{\cal E}}
(f_p + f_{\overline p})\,.
\end{equation}
It is immediately seen that
Eq.\ (\ref{Asmallqb}), or equivalently (\ref{Asmallqc}), reduces
to the results quoted in Eq.\ (\ref{Aex}) when ${\cal Q} = 0$ or
$\Omega = 0$.  However,  Eqs.\ (\ref{Asmallqb}) and (\ref{Asmallqc})
hold also when neither ${\cal Q}$ nor $\Omega$ is zero.
When $q$ is such that $\Omega,{\cal Q}\ll m_e$,
the functions $A,B$ for the electron are given by similar
formulas.

Proceeding in similar form for the functions $C_{n,p}$ and $D_{n,p}$
we obtain
\begin{eqnarray}\label{Csmallq}
C_{n,p} & = & \frac{q^2}{{2\cal Q}^2}\int\frac{d^3{\cal P}}{(2\pi)^32{\cal E}}
\frac{\vec{\cal Q}\cdot\vec v_{\cal P}}
{\Omega - \vec{\cal Q}\cdot\vec v_{\cal P}}\frac{d}{d{\cal E}}
(f_{n,p} - f_{\overline n,\overline p})\\
\label{Dsmallq}
D_{n,p} & = & -\frac{1}{2}\int\frac{d^3{\cal P}}{(2\pi)^3}
\frac{1}{2{\cal E}^2}\left[
\frac{\vec{\cal Q}\cdot\vec v_{\cal P}}{\Omega - \vec{\cal Q}\cdot
\vec v_{\cal P}}
\frac{d}{d{\cal E}}(f_{n,p} + f_{\overline n,\overline p}) 
+ \frac{(f_{n,p} + f_{\overline n,\overline p})}{{\cal E}}
\right]\,.
\end{eqnarray}
Once more, Eqs.\ (\ref{Csmallq}) and (\ref{Dsmallq}) reduce to the
formulas in Eqs.\ (\ref{Cex}) and (\ref{Dex}) in the indicated limits.
We stress also
that, in deriving Eqs. (\ref{Bsmallqb}), (\ref{Asmallqc}),
(\ref{Csmallq}) and (\ref{Dsmallq}), we have made no assumption
regarding the conditions of the background gas.  Thus, 
they are valid for degenerate and non-degenerate
gases, in the relativistic as well the non-relativistic
limits.

\section{Electromagnetic and Neutral-current couplings of the nucleons}

The couplings of the interaction Lagrangian that are
relevant to our calculation are given by
\begin{equation}\label{alint}
L_{int} = -|e|A^\mu(-\overline e\gamma_\mu e + J^{(em)}_\mu)
- g_Z Z^\mu[\overline\nu_L\gamma\nu_L + 
\overline e\gamma_\mu(a_e + 
b_e\gamma_5) e + J^{(Z)}_\mu]\,,
\end{equation}
where, in the standard model $a_e$ and $b_e$ are given in Eq.\ (\ref{neutralcouplings})
while, in terms of the quark fields,
\begin{eqnarray}\label{Jem}
J^{(em)}_\mu & = & \overline q\gamma_\mu \frac{\tau_3}{2} q
+ \frac{1}{6}\overline q\gamma_\mu q\,,\\
\label{JZ}
J^{(Z)}_\mu & = &  \overline q\gamma_\mu \frac{\tau_3}{2} q
- \overline q\gamma_\mu\gamma_5 \frac{\tau_3}{2}q 
- 2\sin^2\theta_W J^{(em)}_\mu\,.
\end{eqnarray}
We have introduced the notation  
\begin{equation}\label{qdoublet}
q = \left(\begin{array}{l}
u\\
d
\end{array}
\right)\,,
\end{equation}
and $\tau_{1,2,3}$ stand for the Pauli matrices.

For each nucleon $f = p,n$, the electromagnetic couplings
are defined by writing the matrix element
\begin{equation}\label{Fem}
\langle f(p^\prime)|J^{(em)}_\mu(0)|f(p)\rangle
= \overline u(p^\prime)\left[F_{1f}^{(em)}\gamma_\mu
- i\frac{F_{2f}^{(em)}}{2m}\sigma_{\mu\nu}q^\nu\right]u(p)\,,
\end{equation}
where $q = p - p^\prime$, $m$ is the nucleon mass 
and $u(p)$ is a Dirac spinor.
The form factors are functions of $q^2$ and are such that
\begin{eqnarray}\label{Fem0}
F_{1p}^{(em)}(0) & = & 1\,,\nonumber\\
F_{1n}^{(em)}(0) & = & 0\,,\nonumber\\
F_{2p}^{(em)}(0) & = & 1.79\,,\nonumber\\
F_{2n}^{(em)}(0) & = & -1.71\,.
\end{eqnarray}
In similar fashion, and using the $SU(2)$ symmetry
property of the matrix elements, we can write
\begin{eqnarray}\label{F3def}
\langle f(p^\prime)|\overline q\gamma_\mu\frac{\tau_3}{2}q|f(p)\rangle
& = & I_{3f}\overline u(p^\prime)\left[F_{1}^{(3)}\gamma_\mu
- i\frac{F_{2}^{(3)}}{2m}\sigma_{\mu\nu}q^\nu\right]u(p)\,,\nonumber\\
\langle f(p^\prime)|\overline q\gamma_\mu\gamma_5\frac{\tau_3}{2}q|f(p)\rangle
& = & I_{3f}\overline u(p^\prime)F_{A}^{(3)}\gamma_\mu\gamma_5 u(p)\,,
\end{eqnarray}
where $I_{3p} = -I_{3n} = 1/2$.  Since $F_A^{(3)}$ is the
same form factor that appears in the charged current
matrix element (which is responsible, for example, for
$\beta$ decay),
\begin{equation}\label{FA0}
g_A \equiv F_A^{(3)}(0) = 1.26\,.
\end{equation}
Further, from the $SU(2)$ decomposition of $J^{(em)}_\mu$
given in Eq.\ (\ref{Jem}), it follows that
\begin{eqnarray}\label{F3}
F_1^{(3)} & = & F_{1p}^{(em)} - F_{1n}^{(em)}\,,\nonumber\\
F_2^{(3)} & = & F_{2p}^{(em)} - F_{2n}^{(em)}\,,
\end{eqnarray}
so that, in particular,
\begin{eqnarray}\label{F30}
F_1^{(3)}(0) & = & 1\,,\nonumber\\
F_2^{(3)}(0) & = & 3.7\,.
\end{eqnarray}

In principle, the form factors that enter the calculation
of the diagram shown in Fig.\ \ref{fig1} are not the
on-shell form factors we have introduced above, but
their off-shell counterpart.  However, since we are
considering situations in which the photon momentum $q$
is small, we will use their value at $q\rightarrow 0$,
for which the formulas given above are valid.
The matrix
element of the neutral-current between nucleon states
can then be written in the form
\begin{equation}\label{jzdef}
\langle f(p^\prime)|J^{(Z)}_\mu(0)|f(p)\rangle
= \overline u(p^\prime)j^{(Z)}_{f\mu}(q)u(p)\,,
\end{equation}
where
\begin{equation}\label{jznucl}
j^{(Z)}_{f\mu}(q) = a_f\gamma_\mu + b_f\gamma_\mu\gamma_5
- i\frac{c_f}{2m}\sigma_{\mu\nu}q^\nu\,.
\end{equation}
From the decomposition of $J^{(Z)}_\mu$ given in Eq.\ (\ref{JZ})
together with Eqs.\ (\ref{Fem}) and (\ref{F3def}), it follows that
\begin{eqnarray}\label{abcnucl}
a_f & = & I_{3f} - 2\sin^2\theta_W Q_f\,,\nonumber\\
b_f & = & -I_{3f}g_A\,,\nonumber\\
c_f & = & I_{3f}[F^{(em)}_{2p}(0) - F^{(em)}_{2n}(0)] - 
2\sin^2\theta_W F^{(em)}_{2f}(0)\,,
\end{eqnarray}
where $Q_p = 1, Q_n = 0$ and, 
as remarked above, the limit $q\rightarrow 0$ is implied.
Similar considerations apply to the electromagnetic vertices
adopted in Eq.\ (\ref{jnucl}).

In Eq.\ (\ref{tmunu}) we have neglected the $c_f$ term in the
nucleon neutral-current couplings
because it appears to be unimportant in the situations
of interest.  However, for completenesss, we summarize
below the results of including such term in the calculation.

The effect of including the $c_f$ term in the definition
of the nucleon neutral-current vertex is taken into
account by making the substitution
\begin{equation}
\gamma_\mu(a_f + b_f\gamma_5) \rightarrow j^{(Z)}_{f\mu}(-q)
\end{equation}
in Eq.\ (\ref{tmunu}).  The end result of making
that substitution is that  $T^{(p,n)}_{\mu\nu}$ can
still be decomposed as in Eq.\ (\ref{tmunundecomp}) but, instead of the formulas
in Eqs.\ (\ref{Tneutron}) and (\ref{Tprot}) for $T^{(p,n}_{T,L}$, we have instead
\begin{eqnarray}
T^{(p)}_T & = & 2\sqrt{2}G_F a_p\left\{
|e|\left(A_{p} - \frac{B_{p}}{{\tilde v}^2}\right) + 
2m_p \kappa_p a_p q^2 D_{p}\right\}
+ \sqrt{2}G_F c_p\left[2|e| q^2 D_p + 
\frac{\kappa_p}{m_p}\left(A^\prime_p - \frac{B^\prime_p}{\tilde v^2}
\right)\right]\,,\nonumber\\
T^{(p)}_L & = & 4\sqrt{2}G_F a_p\left\{
|e|\frac{B_{p}}{{\tilde v}^2} +
m_p \kappa_p a_p q^2 D_{p}\right\}
+ 2\sqrt{2}G_F c_p\left[|e| q^2 D_p + 
\frac{\kappa_p}{m_p}\frac{B^\prime_p}{\tilde v^2}\right]
\end{eqnarray}
and
\begin{eqnarray}
T^{(n)}_T & = & \sqrt{2}G_Fm_n \kappa_n\left[
4a_n q^2 D_n + \frac{c_n}{m^2_n}
\left(A^\prime_n - \frac{B^\prime_n}{\tilde v^2}
\right)\right]\,,\nonumber\\
T^{(n)}_L & = & 2\sqrt{2}G_Fm_n \kappa_n\left[
2a_n q^2 D_n + \frac{c_n}{m^2_n}\frac{B^\prime_n}{\tilde v^2}\right]\,.
\end{eqnarray}
The functions $A^\prime_f, B^\prime_f$ are given by
\begin{eqnarray}
A^\prime_f & = & q^2(\frac{1}{2}A_f + 3m^2_f D_f)\,,\nonumber\\
B^\prime_f & = & -q^2 B_f - {\cal Q}^2(\frac{1}{2}A_f + m^2_f D_f)\,.
\end{eqnarray}
As expected, the additional terms associated with the $c_f$ neutral-current
couplings are accompanied by additional factors of ${\cal Q}$
or $\Omega$ and therefore are not important in the
limit of small photon momentum that we have considered.

\end{document}